\documentclass[12pt]{article}

\usepackage{amssymb}
\usepackage{amsmath}
\usepackage{amscd}
\usepackage{latexsym}
\usepackage{graphicx}
\usepackage{color}

\newcommand{\be}{\begin{equation}}
\newcommand{\ee}{\end{equation}}
\newcommand{\Dlt}{\Delta}
\newcommand{\dlt}{\delta}

\newcommand{\bt}{\beta}
\newcommand{\vp}{\varphi}

\newcommand{\al}{\alpha}
\newcommand{\ra}{\rightarrow}

\newcommand{\gm}{\gamma}

\newcommand{\lbd}{\lambda}
\newcommand{\Lbd}{\Lambda}

\newcommand{\cH}{{\cal H}}

\newcommand{\rgl}{\rangle}
\newcommand{\lgl}{\langle}

\begin{document}

\begin{center}

{\Large{\bf Inconclusive quantum measurements and decisions
under uncertainty} \\ [5mm]

V.I. Yukalov$^{1,2,*}$ and D. Sornette$^{1,3}$} \\ [3mm]

{\it
$^1$ETH Z\"urich\\
Department of Management, Technology and Economics \\
Z\"urich, CH-8092, Switzerland \\ [3mm]

$^2$Bogolubov Laboratory of Theoretical Physics, \\
Joint Institute for Nuclear Research, Dubna 141980, Russia \\ [3mm]

$^3$Swiss Finance Institute, c/o University of Geneva, \\
40 blvd. Du Pont d'Arve, CH 1211 Geneva 4, Switzerland}

\end{center}

\vskip 1cm

\begin{abstract}

We give a mathematical definition for the notion of inconclusive quantum measurements.
In physics, such measurements occur at intermediate stages of a complex measurement
procedure, with the final measurement result being operationally testable. Since the
mathematical structure of Quantum Decision Theory has been developed in analogy with
the theory of quantum measurements, the inconclusive quantum measurements correspond,
in Quantum Decision Theory, to intermediate stages of decision making in the process
of taking decisions under uncertainty. The general form of the quantum probability
for a composite event is the sum of a utility factor, describing a rational evaluation
of the considered prospect, and of an attraction factor, characterizing irrational,
subconscious attitudes of the decision maker. Despite the involved irrationality, the
probability of prospects can be evaluated. This is equivalent to the possibility of
calculating quantum probabilities without specifying hidden variables. We formulate
a general way of evaluation, based on the use of non-informative priors. As an example,
we suggest the explanation of the decoy effect. Our quantitative predictions are in
very good agreement with experimental data.

\end{abstract}

\vskip 3mm

{\parindent=0pt
{\bf Keywords}: quantum measurements, decision theory, inconclusive events,
quantum probability, non-informative priors, decoy effect

\vskip 1cm

$^*$Correspondence: V.I. Yukalov \\
Department of Management, Technology and Economics, \\
ETH Z\"urich (Swiss Federal Institute of Technology at Z\"urich), \\
Scheuchzerstrasse 7,  Z\"urich CH-8092, Switzerland \\
E-mail: yukalov@theor.jinr.ru \\
}

\newpage

\section{Introduction}

The standard theory of quantum measurements (von Neumann 1955) is based on the projection
operator measure corresponding to operationally testable events. Simple measurements really
have to be operationally testable in order to possess physical meaning. However if a
measurement is composite, consisting of several parts, the intermediate stages do not have
to necessarily be operationally testable, but can be inconclusive.

As a typical example, we can recall the known double-slit experiment, when particles pass through
a screen with two slits and then are registered by particle detectors some distance away from the
screen. This experiment can be treated as a composite event consisting of two parts, one being
the passage through one of the slits and second, registration by detectors. The registration of
a particle by a detector is an operationally testable event, since the particle is either detected
or not, with the result being evident for the observer. But the passage of the particle through one
of the slits is not directly observed, and the experimentalist does not know which of the slits the
particle has passed through. In that sense, the passage of the particle through a slit is an
inconclusive event. The existence of this inconclusive event, occurring at the intermediate stage
of the experiment, is intimately associated with an interference effect. Otherwise, if the
experimentalist would precisely determine the slit through which the particle has passed, the
interference pattern registered by the particle detectors would be destroyed. The existence of
interference is precisely due  to the presence of the inconclusive event that happened at the
intermediate stage.

The occurrence of inconclusive events in decision making is even more frequent and important.
Practically any decision, before it is explicitly formulated, passes through a stage of
deliberation and hesitation accompanying the choice. That is, any decision is actually a composite
event consisting of an intermediate stage of deliberation and of the final stage of taking a decision.
The final stage of decision making is equivalent to an operationally testable event in quantum
measurements. While the intermediate stage of deliberation is analogous to an inconclusive event.

The analogy between the theory of quantum measurements and decision theory has been mentioned by
von Neumann (1955). Following this analogy, Quantum Decision Theory (QDT) has been advanced
(Yukalov and Sornette, 2008,2009a,2010,2011,2014a,2015a), with the mathematical structure that
is applicable to both decision making as well as to quantum measurements. The generality of our
framework, being equally suitable for quantum measurements and decision making, is its principal
difference from all other attempts that employ quantum techniques in psychological sciences.
An extensive literature on various quantum models in psychology and cognitive science can be found
in the books (Khrennikov, 2010; Busemeyer and Bruza, 2012; Bagarello, 2013; Haven and
Khrennikov, 2013) and review articles (Yukalov and Sornette, 2009b; Sornette, 2014; Ashtiani and
Azgomi, 2015; Haven and Khrennikov, 2016).

Any approach, applying quantum techniques to decision theory, is naturally based on the notion
of probability. This is because quantum theory is intrinsically probabilistic. Respectively,
the intrinsically probabilistic nature of quantum decision theory is what makes it principally
different from stochastic decision theory, where the choice is treated as always being
deterministic, while in the process of choosing the decision maker acts with errors
(Hey, 1995; Loomes and Sugden, 1998; Carbone and Hey, 2000; Hey, 2005; Conte et al., 2010). Such
stochastic decision theories can be termed as ``deterministic theories embedded into an environment with
stochastic noise". The standard way of using a stochastic approach is to assume a probability
distribution over the values characterizing the errors made by the subjects in the process of
decision making. Then the parameters entering the distribution are fitted to a posteriori empirical
data by maximizing the log-likelihood function. Such a procedure allows one to better fit the given
set of data to the assumed basic deterministic decision theory, in particular due to the introduction
of additional fitting parameters. However, it does not essentially change the structure of the
underlying deterministic theory, although improving it slightly.  And, being descriptive, the
classical stochastic approach does not provide quantitative predictions.

Contrary to classical stochastic theory, in the quantum approach, we do not assume that the choice
of a decision maker is deterministic, with just some weak disturbance by errors. Following the general
quantum interpretation, we consider the choice process, including deliberations, hesitations, and
subconscious estimation of competing outcomes, as intrinsically random. The probabilistic decision,
in the quantum case, is not just a stochastic decoration of a deterministic process, but it is an
unavoidable random part of any choice. The existence of the hidden, often irrational subconscious
feelings and deliberations, results in the appearance of quantum interference and entanglement.
The difference between classical stochastic decision theory and quantum decision theory is similar
to the difference between classical statistical physics and quantum theory. In the former, all
processes are assumed to be deterministic, with statistics coming into play because of measurement
uncertainties, such as no precise knowledge of initial conditions and the impossibility of
measuring exactly the locations and velocities of all particles. In contrast, quantum theory is
principally probabilistic, which becomes especially important for composite measurements.

A detailed mathematical theory of quantum measurements in the case of composite events has been developed
in our previous papers (Yukalov and Sornette, 2013, 2015b, 2016). In the present paper, we concentrate
our attention on composite measurements including intermediate inconclusive events and on the application
of this notion for characterizing decision making under risk and uncertainty. The importance of composite
events, including intermediate inconclusive events, in decision theory makes it necessary to pay a special
attention to the correct mathematical formulation of such events and to the description of their properties
allowing for the quantitative evaluation of the corresponding quantum probabilities. We show that,
despite uncertainty accompanying inconclusive events, it is possible to give quantitative evaluations for
quantum probabilities in decision theory, based on non-informative priors. Considering, as an illustration,
the decoy effect, we demonstrate that even the simple non-informative priors provide predictions in very
good agreement with experimental data.

\section{Composite quantum measurements and events}

In this section, we give a brief summary of the general scheme for defining quantum probabilities for
composite events. As we have stressed above, in our approach, the mathematics is the same for describing
either quantum measurements or decision making. Therefore, when referring to an event, we can keep in mind
either a fact of measurement or a decision action.

Let $A_n$ be a conclusive operationally testable event labelled by an index $n$. And let $B=\{B_\alpha\}$
be a set of inconclusive events labelled by $\alpha$. Defining the space of events as a Hilbert space
$\mathcal{H}$, we associate with an event $A_n$ a state $|n\rangle$ in this Hilbert space and an event
operator $\hat{P}_n$,
\be
\label{1}
A_n \ra | n \rgl \ra \hat P_n = | n \rgl \lgl n | \;   .
\ee
The event operator for an operationally testable event is a projector.

The set of inconclusive events $B$
generates in the Hilbert space $\mathcal{H}$ the state $|B\rangle$ and the event operator $\hat{P}_B$,
\be
\label{2}
B \ra | B \rgl \ra \hat P_B = | B \rgl \lgl B | \;   ,
\ee
where the state reads
\be
\label{3}
  | B \rgl = \sum_\al b_\al | \al \rgl \;  ,
\ee
with coefficients $b_\alpha$ being random complex numbers. The event operator for an inconclusive event
is not necessarily a projector, but a member of a positive operator-valued measure
(Yukalov and Sornette, 2013, 2015a,b, 2016).

The space of events, in the quantum approach, is the Hilbert space
\be
\label{4}
\cH = \cH_A \bigotimes \cH_B
\ee
that is a tensor product of the spaces
$$
 \cH_A = {\rm span} \{ | n \rgl \} \; ,  \qquad
 \cH_B = {\rm span} \{ | \al \rgl \} \; .
$$
Each decision maker is characterized by an operator $\hat{\rho}$ that can be termed the strategic state
of a decision maker, which, in quantum theory, corresponds to a statistical operator. The pair
$\{ {\mathcal H}, \hat{\rho}\}$, in physics, is named a statistical ensemble, and in decision theory, it is
a decision ensemble.

A composite event is called a prospect and is denoted as
\be
\label{5}
 \pi_n = A_n \bigotimes B \;  .
\ee
A prospect $\pi_n$ generates a state $|\pi_n\rangle$ in the Hilbert space of events ${\mathcal H}$
and a prospect operator $\hat{P}_n$,
\be
\label{6}
 \pi_n \ra | \pi_n \rgl \ra \hat P(\pi_n) = | \pi_n \rgl \lgl \pi_n | \;  ,
\ee
with the prospect state
\be
\label{7}
  | \pi_n \rgl =  | n \rgl \bigotimes | B \rgl =  \sum_\al b_\al | n \al \rgl \; .
\ee
The prospect operator is a member of a positive operator-valued measure, which implies that these
operators satisfy the resolution of unity (Yukalov and Sornette, 2013, 2016). Since they contain
random quantities $b_\alpha$, the corresponding random resolution has to be understood not as a
direct equality between numbers, but, e.g., as the equality in mean (Kallenberg, 2001).

The prospect probability is
\be
\label{8}
 p(\pi_n) = {\rm Tr} \; \hat\rho \hat P(\pi_n) \;  ,
\ee
with the trace over the space ${\mathcal H}$. To form a probability measure, the prospect
probabilities are to be normalized:
\be
\label{9}
  \sum_n p(\pi_n) = 1 \; , \qquad 0 \leq p(\pi_n) \leq 1 \;  .
\ee

Taking explicitly the trace in expression (\ref{8}) and separating diagonal and off-diagonal terms,
we see that the prospect probability
\be
\label{10}
p(\pi_n) = f(\pi_n) + q(\pi_n)
\ee
is represented as a sum of a positive-definite term
\be
\label{11}
f(\pi_n) = \sum_\al | b_\al |^2 \lgl  n \al | \hat\rho | n\al \rgl
\ee
and a sign-undefined term
\be
\label{12}
q(\pi_n) = \sum_{\al\neq \bt} b^*_\al b_\al \lgl  n \al | \hat\rho | n\bt \rgl \;  .
\ee

The appearance of the sign-undefined term is a purely quantum effect responsible, in quantum measurements,
for interference patterns. The attenuation of this quantum term is called decoherence. In quantum theory,
decoherence can be due to external as well as to internal perturbations and the influence of measurements
(Yukalov, 2011, 2012a,b). And in quantum decision theory, decoherence can occur due to the accumulation of
information (Yukalov and Sornette, 2015c).

The disappearance of the quantum term implies the transition to classical theory. This is formulated as the
{\it quantum-classical correspondence principle} (Bohr, 1976), which in our case reads as
\be
\label{13}
p(\pi_n) \ra f(\pi_n) \; , \qquad q(\pi_n) \ra 0 \;   .
\ee
This principle tells us that the term $f(\pi_n)$ plays the role of classical probability, hence is to be
normalized:
\be
\label{14}
 \sum_n f(\pi_n) = 1 \; , \qquad 0 \leq f(\pi_n) \leq 1 \;  .
\ee
When decisions concern a choice between lotteries, the classical term $f(\pi_n)$ has to be defined
according to classical decision theory based either on expected utility theory or on some nonexpected
value functional. This suggests to call $f(\pi_n)$ a {\it utility factor}, since it is defined on
rational grounds and reflects the utility of a choice. The quantum term is caused by the interference
and entanglement effects in quantum theory, which correspond, in decision making, to irrational effects
describing the attractiveness of choice. Therefore it can be called the attraction factor. From
equations (\ref{9}) and (\ref{14}), it follows the {\it alternation law}
\be
\label{15}
\sum_n q(\pi_n) = 0 \; , \qquad -1 \leq q(\pi_n) \leq 1 \;  .
\ee

Note that, in quantum theory, the definition of the composite event in the form of prospect (\ref{5})
is valid for any type of operators, since they are defined on different spaces. No problem with
noncommutativity of operators defined on a common Hilbert space arises. This way makes it possible
to introduce joint quantum probabilities for several measurements (Yukalov and Sornette, 2013, 2016).
Contrary to this, considering operators on the same Hilbert space does not allow one to define joint
probabilities for noncommuting operators. Sometimes, one treats the L\"{u}ders probability of
consecutive measurements as a conditional probability. This, however, is not justified from the
theoretical point of view (Yukalov and Sornette, 2013, 2014b, 2016) and also contradicts experimental
data (Boyer-Kassem et al., 2015a,b). But defining the quantum joint probability according to
expression (\ref{8}) contains no contradictions.

In the present section, the general scheme of QDT is presented. Being limited by the length
of this paper, we cannot go into all mathematical details that have been thoroughly described
in our previous publications. However, we would like to stress that for the purpose of practical
applications, it is not necessary to study all these mathematical peculiarities, but it is
sufficient to employ the final formulas following the prescribed rules. One can use the
formulated rules as given prescriptions, without studying their justification. The main formulas
of this section, which are necessary for the following application, are: the form of the
quantum probability (\ref{10}), the normalization conditions (\ref{9}) and (\ref{14}), and
the alternation law (\ref{15}). More details required for practical application will be given
in the sections below.

\section{Non-informative prior for utility factors}

To make the above scheme applicable to decision theory, it is necessary to specify how one should
quantify the values of the utility factors and attraction factors. Here we show how these values can
be defined as non-informative priors.

Let us consider a set of lotteries $L_n = \{x_i, p_n(x_i): i = 1,2,\ldots, N_n\}$, enumerated by
the index $n = 1,2,\ldots,N_L$, with payoffs $x_i$ and their probabilities $p_n(x_i)$. The related
expected utilities $U(L_n) = \sum_i u(x_i) p_n(x_i)$ can be defined according to the expected utility
theory (von Neumann and Morgenstern, 1953). For the present consideration, the utility functions $u(x)$
do not need to be specified. For instance, they can be taken as linear functions, since this choice
has the advantage of making the utility factors independent from the units measuring the payoffs.

In quantum decision theory, the act of choosing a lottery $L_n$, denoted as $A_n$, together with the
accompanying set of inconclusive events $B$, including the decision-maker hesitations (Yukalov and
Sornette, 2014a, 2014b), compose the prospect (\ref{5}). Depending on whether the expected utilities
are positive on negative, there can be two cases.

If the expected utilities of the considered set of lotteries are all positive (non-negative), such
that
\be
\label{16}
U(L_n) \geq 0 \qquad ( n = 1,2, \ldots, N_L ) \;   ,
\ee
then it is reasonable to require that zero utility corresponds to zero utility factor:
\be
\label{17}
 f(\pi_n) \ra 0 \; , \qquad U(L_n) \ra 0 \;  .
\ee
The case where the utility factor is simply proportional to the related expected utility trivially
obeys this condition (\ref{17}). Taking into account the normalization condition (\ref{14}) gives the
utility factor
\be
\label{18}
 f(\pi_n) = \frac{U(L_n)}{\sum_n U(L_n)} \;  .
\ee

When the expected utilities are negative, which happens in the domain of losses, such that
\be
\label{19}
 U(L_n) < 0 \qquad ( n = 1,2, \ldots, N_L ) \;   ,
\ee
the required condition is that an infinite loss corresponds to zero utility factor:
\be
\label{20}
 f(\pi_n) \ra 0 \; , \qquad | U(L_n) | \ra \infty \;   .
\ee
The simplest way to satisfy this condition (\ref{20}) is that the utility factor is inversely proportional
to the related expected utility. Taking into account the normalization condition, we get
\be
\label{21}
f(\pi_n) = \frac{|U(L_n)|^{-1}}{\sum_n |U(L_n)|^{-1}} \;   .
\ee

The utility-factor forms (\ref{18}) and (\ref{21}) coincide with the choice probabilities in the Luce
choice axiom (Luce, 1959). It is possible to show that generalized forms for the utility factors can
be derived by maximizing a conditional Shannon entropy or from the principle of minimal information
(Yukalov and Sornette, 2009b; Frank, 2009; Batty, 2010).

In the case of positive expected utilities, we consider the information functional, taking into account
the normalization condition (\ref{14}) and the expected log-likelihood $\Lambda$. This functional reads
as
\be
\label{22}
I[\; f(\pi_n)\; ] = \sum_n f(\pi_n) \ln f(\pi_n) \; + \;
\lbd \left [ \sum_n f(\pi_n) -1 \right ] \; + \;
 \al \left [ \sum_n f(\pi_n) \Lbd(\pi_n) - \Lbd \right ] \;  ,
\ee
where
$$
 \Lbd(\pi_n) = - \ln U(L_n) \; , \qquad U(L_n) \geq 0 \;  .
$$
Minimizing functional (\ref{22}) results in the utility factor
\be
\label{23}
f(\pi_n) = \frac{U^\al(L_n)}{\sum_n U^\al(L_n)} \qquad (\al > 0) \;   ,
\ee
in which the positive sign of $\alpha$ is prescribed by the condition that the larger utility implies
the larger factor.

In the case of negative expected utilities, the information functional takes the form
\be
\label{24}
I[\; f(\pi_n)\; ] = \sum_n f(\pi_n) \ln f(\pi_n) \; + \;
\lbd \left [ \sum_n f(\pi_n) -1 \right ] \; + \;
 \gm \left [ \Lbd - \sum_n f(\pi_n) \Lbd(\pi_n) \right ] \;   ,
\ee
where
$$
  \Lbd(\pi_n) = - \ln | U(L_n) | \; , \qquad U(L_n) < 0 \;  .
$$
Then its minimization yields the utility factor
\be
\label{25}
f(\pi_n) = \frac{|U(L_n)|^{-\gm}}{\sum_n |U(L_n)|^{-\gm}} \qquad (\gm > 0) \;   ,
\ee
with the positive sign of $\gamma$ prescribed by the requirement that the larger cost implies the smaller
factor.

The utility factors (\ref{23}) and (\ref{25}) are the examples of power-law distributions that are known
in many applications (Frank, 2009; Batty, 2010; Saichev et al. 2010).

\section{Non-informative prior for attraction factors}

Although the attraction factor characterizes irrational features of decision making, it can be estimated by
invoking non-informative prior assumptions. An important consequence of the latter is the {\it quarter law}
derived earlier (Yukalov and Sornette, 2009b, 2010, 2011). Here we first give the new, probably the
simplest, derivation of the quarter law and, second, we show how this law can be used for estimating the
attraction factors in the case of an arbitrary number of prospects.

Let us consider the sum
\be
\label{26}
\frac{1}{N_L} \; \sum_{n=1}^{N_L} | q(\pi_n) | = \int_0^1 \vp(x) x\; dx
\ee
of the attraction factor moduli, where
\be
\label{27}
\vp(x) \equiv \frac{1}{N_L} \; \sum_{n=1}^{N_L} [\; \dlt(x-q(\pi_n) ) + \dlt(x + q(\pi_n))\; ]
\ee
plays the role of the attraction-factor distribution. The latter is normalized as
\be
\label{28}
\int_{-1}^1 \vp(x)\; dx = 1 \;  ,
\ee
since the attraction factors, in view of condition (\ref{15}), vary in the interval $[-1,1]$.
If $q(\pi_n)$ does not equal zero, then normalization (\ref{28}) is evident. And when $q(\pi_n) = 0$,
then one should use the identity
$$
 \int_0^1 \dlt(x) \; dx  =\frac{1}{2} \;
$$
for the semi-integral of the Dirac function.

The use of a non-informative prior implies that the values of the attraction factor are not known.
A full ignorance is captured by a uniform distribution, which, according to normalization (\ref{28}), gives
\be
\label{29}
\vp(x) = \frac{1}{2} \;   .
\ee
In that case, integral (\ref{26}) results in the {\it quarter law}
\be
\label{30}
 \frac{1}{N_L} \; \sum_{n=1}^{N_L} | q(\pi_n) | = \frac{1}{4} \;  .
\ee

If the prospect lattice $\mathcal{L} = \{\pi_n\}$ consists of $N_L$ prospects, we can always arrange
the corresponding attraction factors in the ascending order, such that
\be
\label{31}
q(\pi_n) > q(\pi_{n+1}) \qquad ( n = 1,2, \ldots, N_L-1 ) \;   .
\ee
We denote the largest attraction factor as
\be
\label{32}
 q_{max} \equiv q(\pi_1) > 0 \;  .
\ee
Given the unknown values of the attraction factors, the non-informative prior assumes that they are
uniformly distributed and at the same time they must obey the ordering constraint (\ref{31}).
Then, the joint cumulative distribution of the attraction factors is given by
$$
{\rm Pr}[q(\pi_1)< \eta_1, q(\pi_2)< \eta_2, ..., q(\pi_{N_L})< \eta_{N_L} | \eta_1 \leq \eta_2 \leq ...
\leq \eta_{N_L}] =
$$
\be
= \int_0^{\eta_1} dx_1  \int_{x_1}^{\eta_2} dx_2  ....  \int_{x_{N_L-1}}^{\eta_{N_L}} dx_{N_L}~,
\ee
where the series $\eta_1 \leq \eta_2 \leq ... \leq \eta_{N_L}$ of inequalities ensure the ordering.
It is then straightforward to show that the average values of the $q(\pi_n)$ are equidistant, i.e.
the difference between any two neighboring factors is on average
\be
\label{33}
 \Dlt \equiv \langle q(\pi_n) \rangle - \langle q(\pi_{n+1}) \rangle = const \qquad
 {\rm (independent~of~} n) \;  .
\ee
Taking their average values as determining their typical values, we omit the
symbol $\langle . \rangle$ representing the average operator and use equation (\ref{33})
to represent the $n$-th attraction factor as
\be
\label{34}
 q(\pi_n) = q_{max} - (n-1)\Dlt \;  .
\ee

With notations (\ref{32}) and (\ref{33}), the alternation condition (\ref{15}) yields
\be
\label{35}
 q_{max} = \frac{N_L-1}{2} \; \Dlt \;  .
\ee
And the quarter law (\ref{30}) leads to the gap
\be
\label{36}
 \Dlt = \frac{N_L}{2\sum_n | N_L +1 - 2n| } \;  .
\ee

If $N_L$ is even, then
$$
\sum_{n=1}^{N_L} | N_L + 1 - 2n | = \frac{N_L^2}{2} \qquad (N_L \; even ) \; ,
$$
while when $N_L$ is odd, then
$$
\sum_{n=1}^{N_L} | N_L + 1 - 2n | = \frac{N_L^2-1}{2} \qquad (N_L \; odd ) \;   .
$$
This allows us to represent gap (\ref{36}) as
\begin{eqnarray}
\label{37}
\Dlt = \left \{ \begin{array}{ll}
\frac{1}{N_L} ~~ & ~~ (N_L \; even) \\
\\
\frac{N_L}{N_L^2-1} ~~ & ~~ (N_L \; odd) \; .
\end{array} \right.
\end{eqnarray}
And for the largest attraction factor, we find
\begin{eqnarray}
\label{38}
q_{max} = \left \{ \begin{array}{ll}
\frac{N_L-1}{2N_L} ~~ & ~~ (N_L \; even) \\
\\
\frac{N_L}{2(N_L+1)} ~~ & ~~ (N_L \; odd) \; .
\end{array} \right.
\end{eqnarray}

The above expressions make it possible to evaluate, on the basis of the non-informative prior, the
whole set
\be
\label{39}
Q_{N_L} \equiv \{ q(\pi_n): \; n = 1,2,\ldots,N_L \}
\ee
of the attraction factors:
\begin{eqnarray}
\label{forqLar}
q(\pi_n)= \left \{ \begin{array}{ll}
{1 \over 2N_L}~\left(N_L-2n+1\right)~~ & ~~ (N_L \; even) \\
\\
{N_L \over 2(N_L^2 -1)}~\left(N_L-2n+1\right)
~~ & ~~ (N_L \; odd) \; .
\end{array} \right.
\end{eqnarray}

For example, in the case of two prospects, we have
$$
\Dlt = \frac{1}{2} \; , \qquad q_{max} = \frac{1}{4} \qquad ( N_L = 2 ) \;   ,
$$
which yields the attraction set
$$
 Q_2 = \left \{ \frac{1}{4} \; , ~ -\;\frac{1}{4}  \right \} \; .
$$
For three prospects, we get
$$
\Dlt = \frac{3}{8} \; , \qquad q_{max} = \frac{3}{8} \qquad ( N_L = 3 ) \;   ,
$$
hence
$$
Q_3 = \left \{ \frac{3}{8} \; , ~ 0 \; , ~ -\;\frac{3}{8}  \right \} \;   .
$$
Similarly, for four prospects, we find
$$
 \Dlt = \frac{1}{4} \; , \qquad q_{max} = \frac{3}{8} \qquad ( N_L = 4 ) \;  ,
$$
with the attraction set
$$
Q_4 = \left \{ \frac{3}{8} \; , ~ \frac{1}{8} \; , ~ -\; \frac{1}{8} \; , ~
-\;\frac{3}{8}  \right \} \;   .
$$
When there are five prospects, then
$$
\Dlt = \frac{5}{24} \; , \qquad q_{max} = \frac{5}{12} \qquad ( N_L = 5 ) \;   ,
$$
from where
$$
Q_5 = \left \{ \frac{5}{12} \; , ~ \frac{5}{24} \; , ~ 0 \; , ~
-\; \frac{5}{24} \; , ~ -\;\frac{5}{12}  \right \} \;   .
$$
Thus, we can evaluate the attraction factors for any number of prospects, obtaining
a kind of a quantized attraction set. In the case of an asymptotically large number $N_L$ of
prospects, we have
\be
\label{40}
\Dlt \simeq \frac{1}{N_L} \; , \qquad q_{max} \simeq \frac{1}{2} \qquad ( N_L \gg 1 ) \; ,
\ee
and
\be
 q(\pi_n) \simeq \frac{1}{2} - \frac{2n - 1}{2N_L} \; .
\ee

The non-informative priors can be employed for predicting the results of decision making.
This makes the principal difference compared with the introduction into expected utility of
adjustment parameters that are fitted post hoc to the given experimental data
(Sinisalchi, 2009).

\section{Quantitative explanation of decoy effect}

We now show how the non-informative priors of the attraction factors can be employed to
explain the decoy effect and for quantitative prediction in decision-making.
Throughout this section, we denote, for simplicity, the objects of choice, say $A$, as
well as the act of choosing an object $A$, by the same letter $A$. As has been emphasized
above, the act of choice under uncertainty is a composite prospect. But, again for
simplicity, we employ the same letter for denoting the action $A$ and the related
prospect (\ref{5}).

The decoy effect was first studied by Huber, Payne and Puto (1982), who called it the
effect of asymmetrically dominated alternatives. Later this effect has been confirmed in
a number of experimental investigations
(Simonson, 1989; Wedell, 1991; Tversky and Simonson, 1993; Ariely and Wallsten, 1995).
The meaning of the decoy effect can be illustrated by the following example. Suppose a buyer
is choosing between two objects, $A$ and $B$. The object $A$ is of better quality, but of
higher price, while the object $B$ is of slightly lower quality, while less expensive.
As far as the functional properties of both objects are not drastically different, but $B$ is
cheaper, the majority of buyers value the object $B$ higher. At this moment, the salesperson
mentions that there is a third object $C$, which is of about the same quality as $A$, but of
a much higher price than $A$. This causes the buyer to reconsider the choice between the objects
$A$ and $B$, while the object $C$, having the same quality as $A$ but being much more expensive,
is of no interest. Choosing now between $A$ and $B$, the majority of buyers prefer the
higher quality but more expensive object $A$. The object $C$, being not a choice alternative,
plays the role of a decoy. Experimental studies confirmed the decoy effect for a variety
of objects: cars, microwave ovens, shoes, computers, bicycles, beer, apartments, mouthwash, etc.
The same decoy effect also exists in the choice of human mates distinguished by attractiveness
and sense of humor (Bateson and Healy, 2005). It is common as well for animals, for instance,
in the choice of female frogs of males with different attraction calls characterized either by
low-frequency and longer duration or by faster call rates (Lea and Ryan, 2015).

The decoy effect contradicts the regularity axiom in decision making telling that if $B$
is preferred to $A$ in the absence of $C$, then this preference has to remain in the presence
of $C$.

In the frame of QDT, the decoy effect is explained as follows. Assume buyers consider an
object $A$, which is of higher quality but more expensive, and an object $B$, which is of moderate
quality but cheaper. Suppose the buyers have evaluated these objects $A$ and $B$, which implies that
the initial values of the objects are described by the utility factors $f(A)$ and $f(B)$. In
experiments, the latter correspond to the fractions of buyers evaluating higher the related object.
When the decoy $C$, of high quality but essentially more expensive, is presented, it attracts the
attention of buyers to the quality characteristic. The role of the decoy is well understood as
attracting the attention of buyers to a particular feature of the considered objects, because of which
the decoy effect is sometimes named the attraction effect (Simonson, 1989). In the present case,
the decoy attracts the buyer attention to quality. The attraction, induced by the decoy, is described
by the attraction factors $q(A)$ and $q(B)$. Hence the probabilities of the related choices are now
$$
p(A) =  f(A) + q(A) \; , \qquad p(B) =  f(B) + q(B) \;   .
$$
Since the quality feature becomes more attractive, $q(A) > q(B)$. According to the non-informative prior,
we can estimate the attraction factors as $q(A) = 1/4$ and $q(B) = - 1/4$.

To be more precise, let us take numerical values from the experiment of Ariely and Wallsten (1995), where
the objects under sale are microwave ovens. The evaluation without a decoy results in $f(A) = 0.4$ and
$f(B) = 0.6$. In the presence of the decoy, we predict that the choice probabilities can be evaluated as
$$
p(A) =  f(A) + 0.25 \; , \qquad p(B) =  f(B) - 0.25 \; .
$$
This gives $p(A) = 0.65$ and $p(B) = 0.35$. The experimental values for the choice between $A$ and
$B$, in the presence of but excluding $C$, correspond to the fractions $p_{exp}(A) = 0.61$ and
$p_{exp}(B) = 0.39$, which is close to the predicted probabilities.

Another example can be taken from the studies of the frog mate choice (Lea and Ryan, 2015), where frog
males have attraction calls differing in either low-frequency sound or call rate. The males with
lower frequency calls are denoted as $A$, while those with high call rate, as $B$. In an experiment with
$80$ frog females, without a decoy, it was found that females evaluate higher the fastest call rate,
so that $f(A) = 0.35$ and $f(B) = 0.65$. In the presence of an inferior decoy, attracting attention to
the low-frequency characteristic, the non-informative prior predicts the probabilities
$$
p(A) =  0.35 + 0.25 = 0.6 \; , \qquad p(B) =  0.65 - 0.25 = 0.4 \;   .
$$
The empirically observed fractions are found to be $p_{exp}(A) = 0.6$ and $p_{exp}(B) = 0.4$, in
remarquable agreement with our predictions.

To make it clear how the decoy effect fits the title of the paper "Inconclusive quantum
measurements and decisions under uncertainty", it is worth extending the comments that have
been mentioned in the Introduction.

Our principal point of view is that decision making, generally, almost always deals with
composite events, since any choice is accompanied by subconscious feelings and irrational
biases. The latter are often difficult to formalize and, even more, their weights usually
are not known and are practically unmeasurable. This is why these subconscious irrational
factors can be treated as what is called inconclusive events. When choosing between several
possibilities, say $A_n$, one actually considers composite prospects, as defined in (\ref{5}).
And the composite nature of choices requires the use of quantum techniques, as has been
explained in our previous paper (Yukalov and Sornette, 2014b). Otherwise, the probabilities
of simple events could be characterized by classical theory. It is the composite nature
of the considered prospects that yields the appearance of the quantum term $q(\pi_n)$
related to interference and coherence effects. In that way, the choice between the objects
in the decoy effect is also a composite prospect, being composed of the choice as such and
accompanying subconscious feelings forming an inconclusive set. This is why the use of
QDT here is necessary and why it gives so good results.

It is admissible to give a schematic picture of the choice in the decoy effect by analogy
with the double-slit experiment in physics, which is mentioned in the Introduction. Thus
making a concrete selection of either an object $A$ or $B$ is the analog of the registration
of the particle by a detector. But before such a selection is done, there exists the
uncertainty of deciding which of the object features are actually more important. These
not precisely defined acts of hesitation play the role of the slits, with the uncertainty
associated with which of them the particle has passed through. When it is known which of
the slits the particle has passed through, then the interference effects in physics
disappear. Similarly, in decision theory, if the values of each object are clearly defined,
there are no hesitations, no interference, and the selection can be based on classical rules.
Such an objective evaluation in the decoy effect happens in the absence of any decoy, when
a decision maker rationally evaluates the features of the given objects, say quality and
price. The appearance of a decoy induces hesitations concerning which of the features
are actually more important. These hesitations before the choice are the analogs of
the uncertainty of which slits will be visited by the traveling particle. The uncertainty
results in the interference and the arising quantum term, whether in the registration of
a particle or in the final choice of a decision  maker.

\section{Discussion}

We have presented a mathematical formulation for the concept of inconclusive quantum measurements
and events. This type of measurements in physics happens at intermediate stages of composite measuring
procedures, while the final measurement stage is operationally testable. In decision making, inconclusive
events correspond to the intermediate stage of deliberations. Invoking non-informative priors, it is
possible to estimate the prospect probabilities, thus, predicting the results of decision making.

Generally, invoking more information on the properties of the attraction factor, it is possible to
define its form more accurately than the value given by non-informative prior. For example, from
condition (\ref{9}) it follows that
$$
 -f(\pi_n) \leq q(\pi_n) \leq 1 - f(\pi_n) \;  .
$$
Hence, for a positive $q(\pi_n)$, we have
$$
0 \leq  q(\pi_n) \leq 1 - f(\pi_n) \;   .
$$
While for a negative $q(\pi_n)$, we get
$$
-f(\pi_n) \leq q(\pi_n) \leq 0 \;   .
$$
Therefore, the attraction factor has to satisfy the limits
$$
 q(\pi_n) \ra + 0 \; , \qquad f(\pi_n) \ra 1 \; ,
$$
$$
 q(\pi_n) \ra - 0 \; , \qquad f(\pi_n) \ra 0 \; .
$$
This suggests that the absolute value of the attraction factor can be modelled by an expression
proportional to
$$
 q(\pi_n) \propto f^\mu(\pi_n) [ 1 - f(\pi_n) ]^\nu \;  ,
$$
with $\mu$ and $\nu$ being positive parameters and the sign defined by the ambiguity and risk
aversion principle (Yukalov and Sornette, 2009b, 2010, 2011, 2014a). More detailed study of such
a form will be given in a separate paper.

But it turns out that even the simple non-informative prior provides us a rather good estimate
allowing for quantitative predictions in decision making. And we have illustrated the approach by
the decoy effect for which the non-informative priors yield quantitative predictions in very good agreement
with empirical data.

In this paper, decision making by separate subjects is considered. We think that the
theory can be generalized by considering societies of decision makers. The exchange of
information in a society should certainly influence the decisions of the society members.
To develop a theory of many agents, it is necessary to generalize the approach by treating
a dynamical model of agents exchanging information. Then, we think, it would be feasible
to describe the behavior of the agents operating in a financial market and taking decisions
about buying or selling shares in the presence of information asymmetry. And it would be
possible to explain the known stylized facts in financial markets, such as, for example,
the fat tails of return distributions and volatility clustering, as well as transient
bubbles and crashes, which are connected with herding behavior. Some first results in
that direction are reported in our previous papers (Yukalov and Sornette 2014a,20015c),
where the role of additional information, received by decision makers, is analyzed and it
is shown that the amount of the additional information essentially influences the value
of the quantum term. Further work on the generalization of the approach towards a dynamical
theory of decison-maker societies is in progress.

\vskip 2mm

{\bf Acknowledgment}. Financial support from the ETH Z\"{u}rich Risk Center is appreciated. We
are greateful for discussions to E.P. Yukalova.

\vskip 2mm

{\bf Conflict of Interest Statement}: The authors declare that the research was
conducted in the absence of any commercial or financial relationships that could
be construed as a potential conflict of interest.

\newpage

{\bf References}

\vskip 2mm
Ariely, D., and Wallsten, T.S. (1995).
Seeking subjective dominance in multidimensional space: an explanation of the asymmetric dominance effect.
{\it Org. Behav. Human Decis. Process.}  63, 223--232.

\vskip 2mm
Ashtiani, M., and Azgomi, M.A. (2015).
A survey of quantum-like approaches to decision making and cognition.
{\it Math. Soc. Sci.} 75, 49--50.

\vskip 2mm
Bagarello, F. (2013).
{\it Quantum Dynamics for Classical Systems}. Hoboken: Wiley.

\vskip 2mm
Bateson, M., and Healy, S.D. (2006).
Comparative evaluation and its implications for mate choice.
{\it Trends Ecol. Evol.} 20, 659--664.

\vskip 2mm
Batty, M. (2010).
Space, scale, and scaling in entropy maximizing.
{\it Geogr. Anal.} 42, 395--421.

\vskip 2mm
Bohr, N. (1976).
{\it Collected Works. The Correspondence Principle}, Vol. 3. Amsterdam: North-Holland.

\vskip 2mm
Boyer-Kassem, T., Duch\^{e}n, S., and Guerci, E. (2015a).
Testing quantum-like models of judgment for question order effects.
arXiv:1501.04901.

\vskip 2mm
Boyer-Kassem, T., Duch\^{e}n, S., and Guerci, E. (2015b).
Quantum-like models cannot account for the conjunction fallacy.
Working paper, Tilburg University.

\vskip 2mm
Busemeyer, J.R., and Bruza, P. (2012).
{\it Quantum Models of Cognition and Decision}. Cambridge: Cambridge University.

\vskip 2mm
Carbone, E., and Hey, J.D. (2000).
Which error story is best?
{\it J. Risk Uncert.} 202, 161--176.

\vskip 2mm
Conte, A., Hey, J.D., and Moffatt, P.G. (2011).
Mixture models of choice under risk.
{\it J. Econometr.} 162, 79--88.

\vskip 2mm
Frank, S.A. (2009).
The common patterns of nature.
{\it J. Evol. Biol.} 22, 1563--1585.

\vskip 2mm
Haven, E., and Khrennikov, A. (2013).
{\it Quantum Social Science}. Cambridge: Cambridge University.

\vskip 2mm
Haven, E., and Khrennikov, A. (2016).
Quantum probability and mathematical modelling of decision making.
{\it Philos. Trans. Roy. Soc. A} 374: 20150105.

\vskip 2mm
Hey, J.D. (1995).
Experimental investigations of errors in decision making under risk.
{\it Eur. Econ. Rev.} 39, 633--640.

\vskip 2mm
Hey, J.D. (2005).
Why we should not be silent about noise.
{\it Exp. Econ.} 8, 325--345.

\vskip 2mm
Huber, J., Payne, J.W., and Puto, C. (1982).
Adding asymmetrically dominated alternatives: violations of regularity and similarity hypothesis.
{\it J. Consum. Res.} 9, 90--98.

\vskip 2mm
Kallenberg, O. (2001).
{\it Foundations of Modern Probability}. Berlin: Springer.

\vskip 2mm
Khrennikov, A. (2010).
{\it Ubiquitous Quantum Structure}. Berlin: Springer.

\vskip 2mm
Lea, A.M., and Ryan, M.J. (2015).
Irrationality in mate choice revealed by tungara frogs.
{\it Science} 349, 964--966.

\vskip 2mm
Loomes, G., and Sugden, R. (1998).
Testing different stochastic specifications of risky choice.
{\it Economica} 65, 581--598.

\vskip 2mm
Luce, R.D. (1959).
{\it Individual Choice Behavior: A Theoretical Analysis}. New York: Wiley.

\vskip 2mm
Saichev, A., Malevergne, Y., and Sornette, D. (2010).
{\it Theory of Zipf's Law and Beyond}. Heidelberg: Springer.

\vskip 2mm
Simonson, I. (1989).
Choice based on reasons: the case of attraction and compromise effects.
{\it J. Consum. Res.} 16, 158--174.

\vskip 2mm
Sinisalchi, M. (2009).
Vector expected utility and attitudes toward variation.
{\it Econometrica} 77, 801--855.

\vskip 2mm
Sornette, D. (2014).
Physics and financial economics (1776--2014): puzzles, Ising and agent-based models.
{\it Rep. Prog. Phys.} 77: 062001.

\vskip 2mm
Tversky, A., and Simonson, I. (1993).
Context-dependent preferences.
{\it Manag. Sci.} 39, 1179--1189.

\vskip 2mm
von Neumann, J. (1955).
{\it Mathematical Foundations of Quantum Mechanics}.
Princeton: Princeton University.

\vskip 2mm
von Neumann, J., and Morgenstern, O. (1953).
{\it Theory of Games and Economic Behavior}. Princeton: Princeton University.

\vskip 2mm
Wedell, D.H. (1991).
Distinguishing among models of contextuality induced preference reversals.
{\it Learn. Memory Cogn.} 17, 767--778.

\vskip 2mm
Yukalov, V.I. (2011)
Equilibration and thermalization in finite quantum systems.
{\it Laser Phys. Lett.} 8, 485--507.

\vskip 2mm
Yukalov, V.I. (2012a)
Equilibration of quasi-isolated quantum systems.
{\it Phys. Lett, A} 376, 550--554.

\vskip 2mm
Yukalov, V.I., (2012b).
Decoherence and equilibration under nondestructive measurements.
{\it Ann. Phys. (N.Y.)} 327, 253--263.

\vskip 2mm
Yukalov, V.I., and Sornette, D. (2008).
Quantum decision theory as quantum theory of measurement.
{\it Phys. Lett. A} 372, 6867--6871.

\vskip 2mm
Yukalov, V.I., and Sornette, D. (2009a).
Physics of risk and uncertainty in quantum decision making.
{\it Eur. Phys. J. B} 71, 533--548.

\vskip 2mm
Yukalov, V.I., and Sornette, D. (2009b).
Processing information in quantum decision theory.
{\it Entropy} 11, 1073--1120.

\vskip 2mm
Yukalov, V.I., and Sornette, D. (2010).
Mathematical structure of quantum decision theory.
{\it Adv. Complex Syst.} 13, 659--696.

\vskip 2mm
Yukalov, V.I., and Sornette, D. (2011).
Decision theory with prospect interference and entanglement.
{\it Theor. Decis.} 70, 283--328.

\vskip 2mm
Yukalov, V.I., and Sornette, D. (2013).
Quantum probabilities of composite events in quantum measurements with multimode states.
{\it Laser Phys.} 23: 105502.

\vskip 2mm
Yukalov, V.I., and Sornette, D. (2014a).
Manipulating decision making of typical agents.
{\it IEEE Trans. Syst. Man Cybern. Syst.} 44, 1155--1168.

\vskip 2mm
Yukalov, V.I., and Sornette, D., (2014b).
Conditions for quantum interference in cognitive sciences.
{\it Top. Cogn. Sci.} 6, 79--90.

\vskip 2mm
Yukalov, V.I., and Sornette, D. (2015a).
Positive operator-valued measures in quantum decision theory.
{\it  Lect. Notes Comput. Sci.} 8951, 146--161.

\vskip 2mm
Yukalov, V.I., and Sornette, D. (2015b).
Quantum theory of measurements as quantum decision theory.
{\it J. Phys. Conf. Ser.} 594: 012048.

\vskip 2mm
Yukalov, V.I., and Sornette, D. (2015c).
Role of information in decision making of social agents.
{\it Int. J. Inf. Technol. Dec. Mak.} 14, 1129--1166.

\vskip 2mm
Yukalov, V.I., and Sornette, D. (2016).
Quantum probability and quantum decision making.
{\it Philos. Trans. Roy. Soc. A} 374: 20150100.

\end{document}